\documentclass{JAC2001}
\usepackage{graphicx}
\newcommand{\ud}{\mathrm{d}}
\begin{document}
\title{SPACE-CHARGE DOMINATED BEAM TRANSPORT VIA MULTIRESOLUTION}
\author{Antonina N. Fedorova, Michael G. Zeitlin \\
IPME, RAS, St.~Petersburg, 
V.O. Bolshoj pr., 61, 199178, Russia
\thanks{e-mail: zeitlin@math.ipme.ru}\thanks{http://www.ipme.ru/zeitlin.html;
http://www.ipme.nw.ru/zeitlin.html}}

\maketitle

\begin{abstract}
We consider space-charge dominated beam transport                              
systems, where space-charge forces are the same order as                        
external focusing forces and dynamics of the corresponding emittance growth.       
We consider the coherent modes of oscillations and coherent instabilities          
both in the different nonlinear envelope models                                
and in initial collective dynamics picture described by Vlasov                  
system. Our calculations are based on                                          
variation approach and multiresolution in the base of high-localized            
generalized coherent states/wavelets.                                                   
We control contributions to dynamical processes from underlying           
multiscales via nonlinear high-localized eigenmodes expansions in the         
base of compactly supported wavelet and wavelet packets bases.                 
\end{abstract}

\section{INTRODUCTION}

In this paper we consider the applications of a new nu\-me\-ri\-cal\--analytical 
technique based on wavelet analysis approach for 
calculations related to description of different space-charge effects.
We consider models for space-charge dominated beam transport systems
in case when space-charge forces are the same order as external focusing forces 
and dynamics of the corresponding emittance growth related with oscillations of underlying
coherent modes [1],[2].
Such approach may be useful in all models in which  it is 
possible and reasonable to reduce all complicated problems related with 
collective behaviour and corresponding statistical distributions to the problems described 
by systems of nonlinear ordinary/partial differential 
equations. 
Also we  consider an approach based on 
the second moments of the distribution functions for  the calculation
of evolution of rms envelope of a beam.
The rational type of
nonlinearities allows us to use our results from 
[3]-[14], which are based on the application of wavelet analysis technique to 
variational formulation of initial nonlinear problems.
Wavelet analysis is a set of mathematical
methods, which gives us a possibility to work with well-localized bases in
functional spaces and give for the general type of operators (differential,
integral, pseudodifferential) in such bases the maximum sparse forms. 
In part 2 we describe the approach based on Vlasov-type model and corresponding rms equations.
In part 3 we present explicit analytical construction for solutions of
Vlasov (besides gaussians) and rms equations from part 2, which are based on variational formulation 
of initial dynamical problems, multiresolution representation and fast wavelet transform technique [15].
We give explicit representation for all dynamical variables in the base of
high-localized generalized coherent states/wavelets. 
Our solutions are parametrized
by solutions of a number of reduced algebraical problems
one from which 
is nonlinear with the same degree of nonlinearity and the others  are
the linear problems which come from the corresponding 
wavelet constructions.
In part 4 we present results of numerical calculations.

\section{VLASOV/RMS EQUATIONS}

Let 
${\bf x}(s)=\big(x_1(s),x_2(s)\big)$
be the transverse coordinates, then single-particle equation of motion is
(ref.[1] for designations):
\begin{eqnarray}
{\bf x}''+\mathbf{k}_{\bf x}(s){\bf x}-D\mathbf{E}_\mathbf{x}({\bf x},s)=0,
\end{eqnarray}
where $D=q/{m\gamma^3 v_0^2}$,  $\ \mathbf{k}_{\mathbf{x}}(s)$ describes periodic focusing force and 
$\mathbf{E}$ satisfies the Poisson equation:
\begin{equation}
\nabla\cdot{\bf E}=\frac{q}{\varepsilon_0}n({\bf x},s)
\end{equation}
with the following density projection:
\begin{equation}
n=\int\int f({\bf x},{\bf x}',s) \ud {\bf x}'
\end{equation}
Distribution function $f(\mathbf{x},\mathbf{x}')$  satisfies Vlasov equation
\begin{equation}
\frac{\partial f}{\partial s}+({\bf x}'\cdot\nabla) f - \Big({\bf k}-
D{\bf E}\Big)\cdot\nabla_{\bf x'} f=0
\end{equation}
Using standard procedure, which takes into account that rms
emittance is described by the seconds moments only
$$
\varepsilon^2_{x,rms}=<x_i^2><x_j^2>-<x_i x_j>^2 (i\ne j),
$$
we arrive to the beam envelope equations for $\sigma_{x_i}$:
{\setlength\arraycolsep{1pt}
\begin{eqnarray}
\sigma_{x_1}''+k_{x_1}(s)\sigma_{x_1}-\varepsilon^2_{x_1}/\sigma_{x_1}^3-C/(\sigma_{x_1}+\sigma_{x_2})&=&0 \\
\sigma_{x_2}''+k_{x_2}(s)\sigma_{x_2}-\varepsilon^2_{x_2}/\sigma_{x_2}^3-C/(\sigma_{x_1}+\sigma_{x_2})&=&0, \nonumber
\end{eqnarray}}
where $C=qI/\pi\varepsilon_0 m\gamma^3 v_0^2$
but only in case when we can calculate explicitly $<x_iE_j>$.
An additional equation describes evolution of 
$\varepsilon^2_{x_i}(s)$:
\begin{equation}
\frac{\ud\varepsilon_x^2}{\ud s}=32D\big(<x_i^2><x_jE_i>-<x_ix_j><x_iE_i>\big)\nonumber
\end{equation}
For nonlinear $E_i$ we need  higher order moments, which lead  to infinite system of equations.
These rms-type envelope equations,
from the formal  point of view, 
are not more than nonlinear differential equations with rational
nonlinearities and variable coefficients.

\section{WAVELET REPRESENTATIONS}

One of the key points of wavelet approach demonstrates that for a large class of
operators wavelets are good approximation for true eigenvectors and the corresponding 
matrices are almost diagonal. Fast wavelet transform gives  the maximum sparse form of operators
under consideration.
It is true also in case of our Vlasov-type system of equations (1)-(4). We have both differential 
and integral operators inside.
So, let us denote our (integral/differential) operator from equations (1)-(4)  as  $T$ and his kernel as $K$.
We have the following representation:
\begin{equation}
<Tf,g>=\int\int K(x,y)f(y)g(x)\ud x\ud y
\end{equation}
In case when $f$ and $g$ are wavelets $\varphi_{j,k}=2^{j/2}\varphi(2^jx-k)$
(7) provides the standard representation of operator $T$.
Let us consider multiresolution representation
$$
\dots\subset V_2\subset V_1\subset V_0\subset V_{-1}
\subset V_{-2}\dots
$$ 
The basis in each $V_j$ is 
$\varphi_{j,k}(x)$,
where indices $\ k, j$ represent translations and scaling 
respectively or the action of underlying affine group
which act as a ``microscope'' and allow us to construct
corresponding solution with needed level of resolution.
Let $T$ act : $L^2(R^n)$
$\rightarrow L^2(R^n)$, with the kernel $K(x,y)$ and
$P_j: L^2(R^n)\rightarrow V_j$ $(j\in Z)$ is projection
operators on the subspace $V_j$ corresponding to j level of resolution:
$$
(P_jf)(x)=\sum_k<f,\varphi_{j,k}>\varphi_{j,k}(x).
$$ 
Let
$Q_j=P_{j-1}-P_j$ be the projection operator on the subspace $W_j$ then
we have the following "microscopic or telescopic"
representation of operator T which takes into account contributions from
each level of resolution from different scales starting with
coarsest and ending to finest scales [15]:
$$
T=\sum_{j\in Z}(Q_jTQ_j+Q_jTP_j+P_jTQ_j).
$$
We remember that this is a result of presence of affine group inside this
construction.
The non-standard form of operator representation [15] is a representation of
operator T as  a chain of triples
$T=\{A_j,B_j,\Gamma_j\}_{j\in Z}$, acting on the subspaces $V_j$ and
$W_j$:
$$
 A_j: W_j\rightarrow W_j, B_j: V_j\rightarrow W_j,
\Gamma_j: W_j\rightarrow V_j,
$$
where operators $\{A_j,B_j,\Gamma_j\}_{j\in Z}$ are defined
as
$A_j=Q_jTQ_j, \quad B_j=Q_jTP_j, \quad\Gamma_j=P_jTQ_j.$
The operator $T$ admits a recursive definition via
$$T_j=
\left(\begin{array}{cc}
A_{j+1} & B_{j+1}\\
\Gamma_{j+1} & T_{j+1}
\end{array}\right),$$
where $T_j=P_jTP_j$ and $T_j$ works on $ V_j: V_j\rightarrow V_j$.
It should be noted that operator $A_j$ describes interaction on the
scale $j$ independently from other scales, operators $B_j,\Gamma_j$
describe interaction between the scale j and all coarser scales,
the operator $T_j$ is an "averaged" version of $T_{j-1}$.
We may compute such non-standard representations of operator
for different operators (including pseudodifferential). As in case of differential operator 
$\ud/\ud x$ as in other cases
in the
wavelet bases we need only to solve the system of linear algebraical
equations. 
Let 
$$
r_\ell=\int\varphi(x-\ell)\frac{\ud}{\ud x}\varphi(x)\ud x, \ell\in Z.
$$
Then, the representation of $d/dx$ is completely determined by the
coefficients $r_\ell$ or by representation of $d/dx$ only on
the subspace $V_0$. 
The coefficients $r_\ell, \ell\in Z$ satisfy the
usual system of linear algebraical equations.
For the representation of operator $d^n/dx^n$ or integral operators we have the similar reduced
linear system of equations.
Then finally we have for action of operator $T_j(T_j:V_j\rightarrow V_j)$
on sufficiently smooth function $f$:
$$
(T_j f)(x)=\sum_{k\in Z}\left(2^{-j}\sum_{\ell}r_\ell f_{j,k-\ell}\right)
\varphi_{j,k}(x),
$$
where $\varphi_{j,k}(x)=2^{-j/2}\varphi(2^{-j}x-k)$ is wavelet basis and
$$
f_{j,k-1}=2^{-j/2}\int f(x)\varphi(2^{-j}x-k+\ell)\ud x
$$
are wavelet coefficients. So, we have simple linear para\-met\-rization of
matrix representation of our operators in wavelet bases
and of the action of
this operator on arbitrary vector in our functional space.
Then we may apply our approach from [3]-[14].
For constructing the solutions of rms type equations (5),(6) obtained from Vlasov equations
we also use our variational approach, which reduces initial problem to the problem of
solution of functional equations at the first stage and some
algebraical problems at the second stage.
We have the solution in a compactly
supported wavelet basis.
Multiresolution representation is the second main part of our construction.
The solution is parameterized by solutions of two reduced algebraical
problems, one is nonlinear and the second are some linear
problems, which are obtained from one of the wavelet
constructions.
The solution of equations (5),(6) has the following form
\begin{equation}\label{eq:z}
z(s)=z_N^{slow}(s)+\sum_{j\geq N}z_j(\omega_js), \quad \omega_j\sim 2^j
\end{equation}
which corresponds to the full multiresolution expansion in all underlying  
scales.
Formula (\ref{eq:z}) gives us expansion into a slow part $z_N^{slow}$
and fast oscillating parts for arbitrary N. So, we may move
from coarse scales of resolution to the 
finest one to  obtain  more detailed information about our dynamical process.
The first term in the RHS of representation (8) corresponds on the global level
of function space decomposition to  resolution space and the second one
to detail space. In this way we give contribution to our full solution
from each scale of resolution or each time scale.
It should be noted that such representations (8)
give the best possible localization
properties in  phase space. This is especially important because 
our dynamical variables correspond  to distribution functions/moments of ensemble of beam particles.
\begin{figure}[htb]
\centering
\includegraphics*[width=60mm]{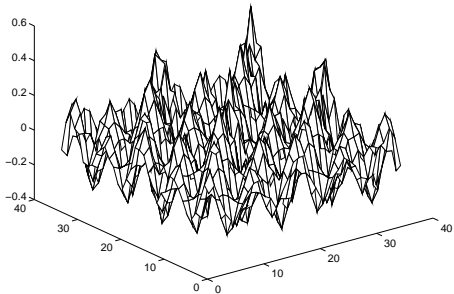}
\caption{6-eigenmodes representation.}
\end{figure} 

\begin{figure}[htb]
\centering
\includegraphics*[width=60mm]{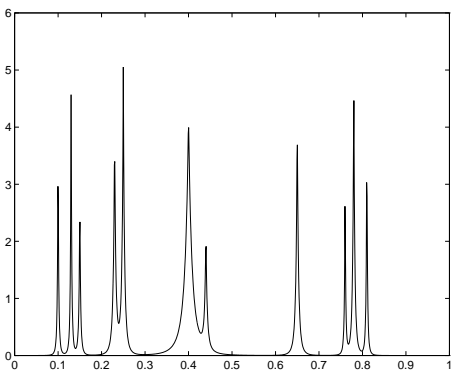}
\caption{Region of nonlinear resonances.}
\end{figure} 

\begin{figure}[htb]
\centering
\includegraphics*[width=60mm]{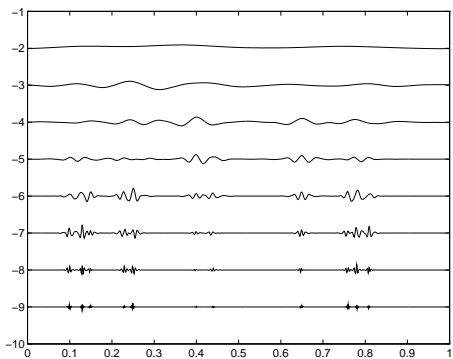}
\caption{Eigenmodes decomposition.}
\end{figure}  
\section{Numerical Calculations}

Now we present numerical illustrations of previous analytical
approach. Numerical calculations are based on compactly supported
wavelets and related wavelet families.
Fig.~1 demonstrates 6-scale/eigenmodes construction for solution of
equations (4). 
Figures~2,3 demonstrate resonances region and corresponding nonlinear coherent eigenmodes decomposition
according to equation (8) [16]. 

\section{ACKNOWLEDGMENTS}

We would like to thank The U.S. Civilian Research \& Development Foundation (CRDF) for
support (Grants TGP-454, 455), which gave us the possibility to present our nine papers during
PAC2001 Conference in Chicago and Ms.Camille de Walder from CRDF for her help and encouragement.


\begin{thebibliography}{16}

\bibitem{1}
I. Hofmann, CERN Proc.95-06, vol.2, 941, 1995

\bibitem{2}
The Physics of High Brightness Beams, Ed.J. Rosenzweig \& L. Serafini,
World Scientific, 2000

\bibitem{3}
A.N. Fedorova and M.G. Zeitlin, 
 {\it Math. and Comp. in Simulation}, {\bf 46}, 527, 1998.

\bibitem{4}
A.N. Fedorova and M.G. Zeitlin,
{\it New Applications of Nonlinear and Chaotic Dynamics in Mechanics}, 31, 101
Klu\-wer,  1998.

\bibitem{5}
A.N. Fedorova and M.G. Zeitlin,
{\bf CP405}, 87, American Institute of Physics, 1997.
Los Alamos preprint,\\
 physics/9710035.

\bibitem{6}
A.N. Fedorova, M.G. Zeitlin and Z.~Parsa, 
Proc. PAC97 
{\bf 2}, 1502, 1505, 1508, APS/IEEE, 1998.

\bibitem{7}
A.N. Fedorova, M.G. Zeitlin and Z.~Parsa, 
Proc. EPAC98, 930, 933, Institute of Physics, 1998.

\bibitem{8}
A.N. Fedorova, M.G. Zeitlin and Z.~Parsa,    
{\bf CP468}, 48, American Institute of Physics, 1999.
Los Alamos preprint, physics/990262.

\bibitem{9}
A.N. Fedorova, M.G. Zeitlin and Z.~Parsa,  
{\bf CP468}, 69, American Institute of Physics, 1999.
Los Alamos preprint, physics/990263.

\bibitem{10}
A.N. Fedorova and M.G. Zeitlin,  
Proc. PAC99, 
1614, 1617, 1620, 2900, 2903,
2906, 2909, 2912, APS/IEEE, New York, 1999.\\
Los Alamos preprints: 
physics/9904039, physics/9904040,\\
 physics/9904041, physics/9904042, physics/9904043, \\
physics/9904045, physics/9904046, physics/9904047.

\bibitem{11}
A.N. Fedorova and M.G. Zeitlin,
The Physics of High Brightness Beams, 235, World Scientific, 2000. 
Los Alamos preprint: physics/0003095.

\bibitem{12}
A.N. Fedorova and M.G. Zeitlin,  Proc. EPAC00, 415, 872,  1101, 1190, 1339, 2325,Austrian Acad.Sci.,2000.\\ 
Los Alamos preprints: physics/0008045, physics/0008046,\\
 physics/0008047, physics/0008048, physics/0008049,\\
 physics/0008050.

\bibitem{13}
A.N. Fedorova, M.G. Zeitlin, Proc. 20 International Linac Conf., 300, 303, SLAC, Stanford, 2000. 
Los Alamos pre\-pri\-nts: physics/0008043, physics/0008200.

\bibitem{14}
A.N. Fedorova, M.G. Zeitlin, Los Alamos preprints:\\
 physics/0101006, physics/0101007
and World Scientific, in press. 

\bibitem{15}
G. Beylkin, R. Coifman, V. Rokhlin, CPAM, 44, 141, 1991

\bibitem{16}
D. Donoho, WaveLab, Stanford, 1998.


\end{thebibliography}
\end{document}